# Spin-orbit Torque and Spin Hall Effect-based Cellular Level Therapeutic Neuromodulators: Modulating Neuron Activities through Spintronic Nanodevices

Kai Wu[1], Diqing Su[2], Renata Saha[1], and Jian-Ping Wang[1, *]

[1]Department of Electrical and Computer Engineering, University of Minnesota, Minneapolis, Minnesota 55455, USA

[2]Department of Chemical Engineering and Material Science, University of Minnesota, Minneapolis, Minnesota 55455, USA

*Corresponding author E-mail: jpwang@umn.edu

**Abstract:** Artificial modulation of a neuronal subset through ion channels activation can initiate firing patterns of an entire neural circuit *in vivo*. As nanovalves in the cell membrane, voltage-gated ion channels can be artificially controlled by the electric field gradient that caused by externally applied time varying magnetic fields. Herein, we theoretically investigate the feasibility of modulating neural activities by using magnetic spintronic nanostructures. An antiferromagnet/ferromagnet (AFM/FM) structure is explored as neuromodulator. For FM layer with perpendicular magnetization, stable bidirectional magnetization switching can be achieved by applying in-plane currents through AFM layer to induce the spin-orbit torque (SOT) due to the spin Hall effect (SHE). This Spin-orbit Torque Neurostimulator (SOTNS) utilizes in-plane charge current pulses to switch the magnetization in FM layer. The time changing magnetic stray field induces electric field that modulates the surrounding neurons. The Object Oriented Micromagnetic Framework (OOMMF) is used to calculate space and time dependent magnetic dynamics of SOTNS structure. The current driven magnetization dynamics in SOTNS has no mechanically moving parts. Furthermore, the size of SOTNS can be down to tens of nanometers, thus, arrays of SOTNSs could be fabricated, integrated together and patterned on a flexible substrate, which gives us much more flexible control of the neuromodulation with cellular resolution.

## 1. Introduction

Nowadays, technologies for controlled artificial modulation of neuronal activities are recognized as important tools for neural science and engineering. The past three decades have seen a tremendous advance in neuromodulation technologies, including noninvasive approaches like Transcranial Magnetic Stimulation (TMS) and invasive strategies such as Deep Brain Stimulation (DBS). Each of these emerging approaches has its unique strengths and applications, but also drawbacks and limitations. DBS is an invasive technique in which electrodes are implanted inside the brain permanently to activate deeply located neurons [1, 2]. The high impedance between the DBS electrodes and neurons and the corrosion of the electrodes significantly decrease the effectiveness of this technique. Furthermore, the implanted electrodes can be affected by the migration of cells (such as astrocytes) that attempt to seal off the devices, causing increased impedance and alterations in the electric field. This is seen in the need to constantly re-program currently available DBS devices. On the other hand, TMS is a non-invasive technique in which the brain's electrical environment is modulated by passing strong non-static magnetic fields (1.5 – 3 T) through scalp and skull. These magnetic fields are generated by passing alternating current (AC) through a coil with ferromagnetic core [3-5]. However, TMS cannot stimulate neurons selectively because it cannot produce a focused magnetic field. In addition to being bulky in size, TMS cannot activate deeply located neurons since the magnetic field decays exponentially over distance.

The quest to miniaturize the TMS devices and meanwhile exploit the efficiency of magnetic neurostimulators, Bonsammer *et.al* have pioneered in neuron stimulations with sub-mm coils [6-8]. They simulated as well as prototyped microcoils and performed experiments on several animal-models, validating microcoils to be a promising alternative for the existing neurostimulators. However, these micro-magnetic stimulating coils (µMS) are capable of effectively stimulating neurons only when the axis of the coils are parallel to the axis of the axons (see Figure 3 of Ref. [8]). This makes these µMS coils quite inefficient in terms of directionality as in reality neurons are randomly oriented. In this work, we are exploring the feasibility of modulating neural activities by using magnetic spintronic nanostructures. To the best of our knowledge, no studies have yet been reported demonstrating the applications of magnetic spintronic nanostructures for any kind of neurostimulations. This Spin-orbit Torque Neurostimulator (SOTNS) utilizes in-plane charge current pulses to flip the magnetization in ferromagnet nanopillar back and forth. As a result, the time varying magnetic stray field induces electric field that modulates the surrounding neurons. Compared to DBS, SOTNS generates an electric field gradient (by means of a time-varying magnetic field) that is not affected by the encapsulation of astrocytes or any other cells. SOTNS is a nanoscale device, capable of stimulating neurons in any orientations as opposed to µMS coils. In addition, the current driven magnetization dynamics in SOTNS has no mechanically moving parts. Furthermore, the ability to fabricate magnetic nanostructures allows neurostimulation at the level of single cells and potentially at the synapse level.

## 2. Spin-orbit Torque Neuromodulator (SOTNS)

**2.1 Physical Model**

Recent studies have shown that by applying an in-plane charge current $J_c$ to a heterostructure with large spin-orbit interaction and structural inversion asymmetry results in a Spin-orbit Torque (SOT), which in turn induces the magnetization switching in the adjacent perpendicular ferromagnetic layer [9]. A typical structure consists of a nonmagnet/ferromagnet (NM/FM) bilayer, where the NM layer is a spin Hall channel and it is responsible to generate transverse spin current $J_s$. This spin current is polarized along the in-plane direction and it accumulates at the interface between NM and FM layers exerting a Slonczewski-like torque (SLT) to switch the magnetization of FM layer. The ratio of spin current to the charge current is represented by the spin Hall angle (SHA). Usually a small symmetry-breaking bias field along the charge current direction is needed for deterministic SOT switching [10]. However, it is impractical to apply an external magnetic field in the neuromodulation applications. Recently, some field-free SOT switching methods have been reported by introducing interlayer exchange coupling [11-13], tilted magnetic anisotropy [14], interlayer dipole coupling [15], etc.

Herein, we are exploring the antiferromagnet/ferromagnet (AFM/FM) structure as neuromodulator (see Figure 1(a) & (b)), where the AFM layer supplies an in-plane exchange bias field on the adjacent FM layer and meanwhile generates SOT that enables the purely electrical deterministic switching of perpendicular magnetization without any external bias field. For a system with single-domain state and perpendicular magnetic anisotropy (PMA) on top of an AFM layer with a current-in-plane (CIP) geometry, the critical current density $J_C$ for a deterministic magnetization switching is written as [16]:

$$J_C = \frac{2e\mu_0}{\hbar} \frac{M_S t_F}{\theta_{SH}^{eff}} \left( \frac{H_K^{eff}}{2} - \frac{H_X}{\sqrt{2}} \right)$$

$$H_K^{eff} = H_K - N_d M_S$$

where $\theta_{SH}^{eff}$ is the effective spin Hall angle, $M_S$, $t_F$, $H_K^{eff}$, and $H_K$ are the saturation magnetization, thickness, effective anisotropy field, and perpendicular anisotropy field of FM layer, respectively, and $H_X$ is the exchange bias field.

One advantage of this SOTNS is that there are no mechanical moving parts in this system, the magnetization (as well as the direction of magnetic field) of FM layer can flip back and forth by changing the charge current directions as shown in Figure 1(a) & (b), where an in-plane charge current $J_c$ passing through the AFM layer is converted into a perpendicular spin current $J_s$ due to the spin Hall effect (SHE).

Another advantage of SOTNS is that the magnetic field is not affected by the encapsulation of astrocytes or any other cells. Thus, the issues such as corrosion of the electrodes and frequently re-programing for DBS devices won't happen for a SOTNS system. Based on the Maxwell-Faraday law, rapidly fluxing magnetic field induces electromotive force (emf):

$$\oint \boldsymbol{E} \cdot d\boldsymbol{l} = -\iint \frac{\partial \boldsymbol{B}}{\partial t} \cdot d\boldsymbol{S}$$

where $\boldsymbol{B}$ is the magnetic field, $\boldsymbol{E}$ is the electric field, $\boldsymbol{l}$ is the contour, and $\boldsymbol{S}$ is the surface.

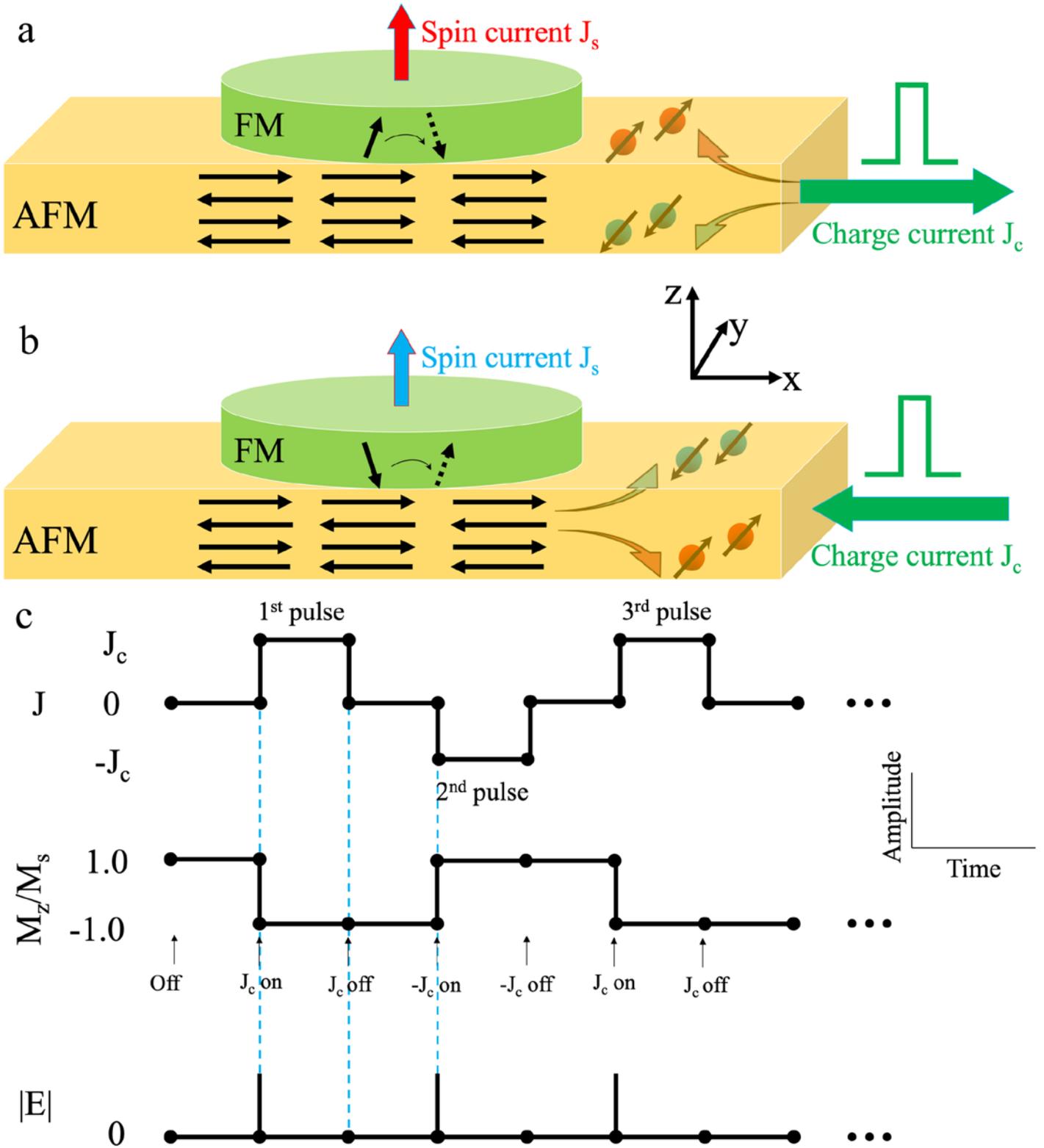

Figure 1. (a) and (b) are Schematics of the antiferromagnet/ferromagnet (AFM/FM) bilayer structure. $J_C$ is the in-plane charge current along X-direction. Exchange bias supplied by AFM allows field-free switching in laterally

homogeneous structures. In-plane charge current running in X-direction yields a spin current in Z-direction due to the spin Hall effect, this spin current exerts torques to switch the magnetization of FM layer. (c) Theoretical switching behavior of SOTNS model.

Figure 1(c) describes the theoretical switching of the magnetization $M_Z$ (the Z-component of magnetization $M$) from +Z to -Z directions and vice versa depending on the charge current direction applied during the time window. Upon the onsets of magnetization $M_Z$ flipping from +Z to -Z directions (or -Z to +Z directions, see the black arrows in Figure 2(b)), the time-varying magnetic flux induces local electric field $E$ (see the red circles in Figure 2(b)). Many systematic explorations have been carried out on the effect of stimulation parameters including stimulation intensity, frequency, and pulse width to establish optimal therapeutic ranges. It is reported that the modulation of neuron activities requires voltage gradients of 5 – 10 mV/mm [17], typical electric field pulse width values vary from 20 μs to 120 μs while higher pulse width values require lower stimulus intensities to achieve a required clinical effect, and the stimulus pulse frequency varies from 60 Hz up to 200 Hz [1].

The size of SOTNS can be down to tens of nanometers, thus, arrays of SOTNSs could be fabricated and integrated together as shown in Figure 2(a). The SOTNSs patterned on a flexible substrate will give us much more flexible control of the neuromodulation. While this proof-of-concept device includes a single chip, dozens of chips could be run by the same electronics. These devices can then be assembled to make a full 3-D array. This is similar in concept to the work by Wise et al. [18, 19], but with spintronic nanodevices and nanofabrication technology, the size of each neuromodulator could shrink from μm to nm.

**2.2 Biological Model**

Neurons encode information with electrical signals and transmit the information to other neurons by synapses. The action potential is a very important electrical signal that occurs when the membrane potential of a specific axon location rapidly rises and falls. Action potentials are generated mainly by specific types of ion channels embedded in a cell's membrane. The principle ions involved in an action potential are $Na^+$ and $K^+$ ions with $Na^+$ ions enter the cell and $K^+$ ions move out of the cell.

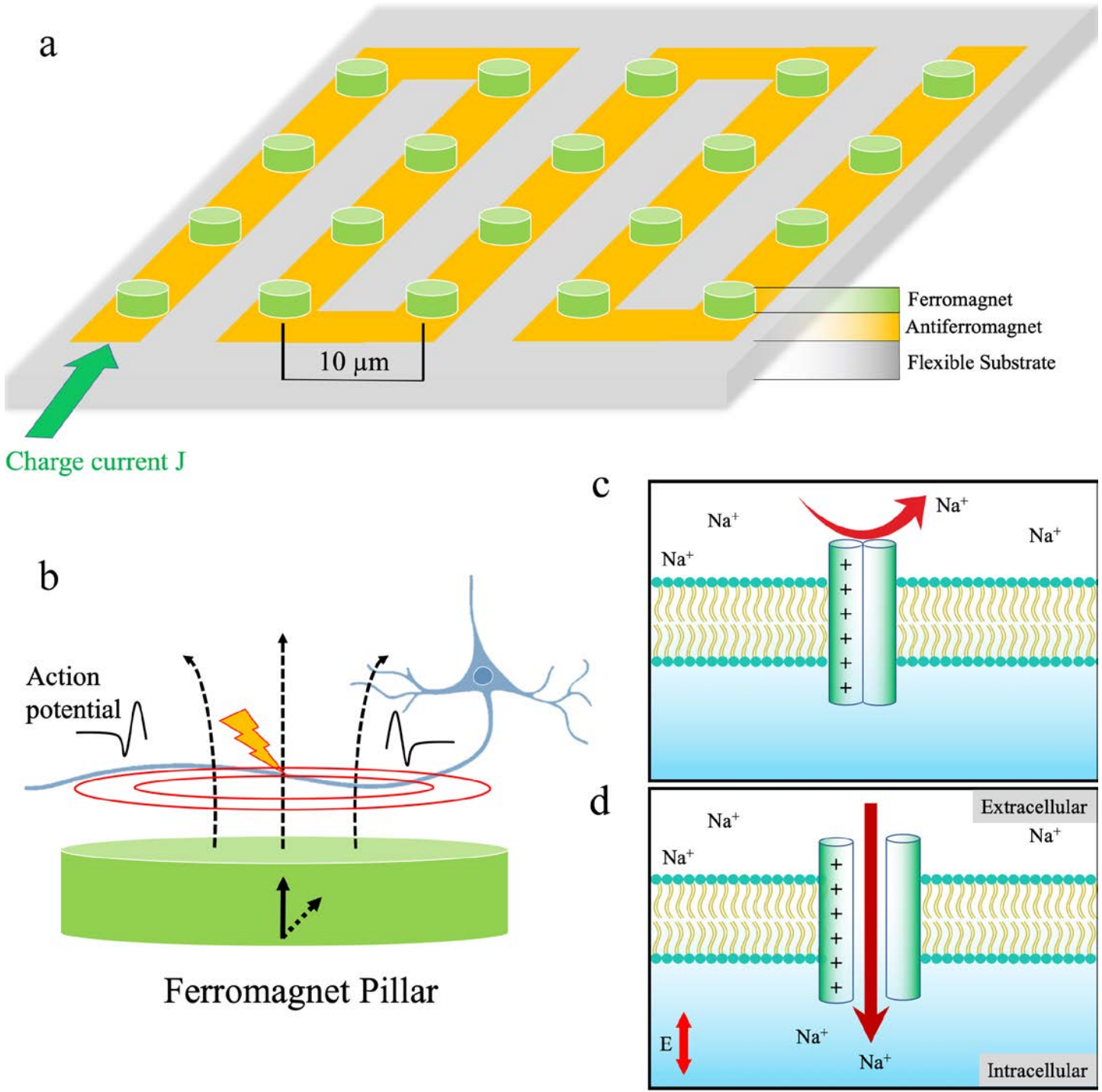

Figure 2. (a) SOTNS arrays integrated on a flexible substrate give us much more flexible control of the neuromodulation. (b) Flipping the magnetization of the FM layer induces electric field that can modulate the neural activities of nearby neurons, initiating two action potentials on the axon going to opposite ways: the orthodromic and antidromic. (c) Schematic view of voltage-gated sodium channel (VGSC) before a stimulation. The Na$^+$ channels are in the deactivated state. (d) The VGSC opens upon the onset of an electric field $E$, allowing Na$^+$ ions to flow into the neuron through the channel. An action potential is initiated.

These ion channels are pore-forming membrane proteins that allow ions to pass through the channel pore. They underlie the resting membrane potential, action potentials, and other electrical signals by gating the flow of ions across the cell membrane. There are three main groups of ion channels, they can be classified by the nature of their gating: voltage-gated, ligand-gated, and mechanically-gated. To understand the neural circuits and behaviors, many external means have been used to activate ion channels, such as the ultrasonic control of neural activities through the mechanically-gated ion channels [20, 21], the chemical neuromodulation exerting on the ligand-gated ion channels that delivers drug invasively (e.g., memantine and agomelatine) for brain disorder treatments [22, 23], the TMS and DBS techniques which are artificially controlling the voltage-gated ion channels. As the ion channels open by different means, specific types of ions pass through the channels down their electrochemical gradient, initiating an action potential. If the stimulation is exerting on the ion channels at the axon hillock region, then one action potential will be initiated and travel down the axon (orthodromic). If the stimulation is exerting in the middle of an axon, two action potentials will be initiated and propagate along opposite directions (antidromic and orthodromic). Figure 2(b) shows the latter case where the electric field generated by SOTNS is exerting in the middle of an axon.

Take the voltage gated sodium channel (VGSC) as an example. As shown in Figure 2(c), before a stimulation, the axonal membrane is at its resting potential and the Na$^+$ channels are in the deactivated state, blocking the Na$^+$ ions in the extracellular side. During one cycle of neuromodulation process, an electric field $\boldsymbol{E}$ is generated by SOTNS locally. In response to the electric field, the VGSCs open (see Figure 2(d)), allowing Na$^+$ ions to flow into the neuron through the channels, causing the voltage across the neuronal membrane to increase. The increase in membrane potential constitutes the rising phase of an action potential.

## 3. Micromagnetic Simulations on SOTNS

In this work, the Object Oriented Micromagnetic Framework (OOMMF) [24] is used to calculate space and time dependent magnetic dynamics of SOTNS structure. For FM layer with perpendicular magnetization, stable bidirectional magnetization switching can be achieved by applying in-plane currents through AFM layer to induce the SOT under an exchange bias field. Many experimental works have revealed that the SLT ($\tau_{SLT}$) is considered to be responsible for the SOT switching [10, 25-28]. The Landau-Lifshitz-Gilbert (LLG) equation including SOT term is expressed as:

$$\frac{\partial \boldsymbol{m}}{\partial t} = -\gamma_0 \boldsymbol{m} \times \boldsymbol{H}_{eff} + \alpha \boldsymbol{m} \times \frac{\partial \boldsymbol{m}}{\partial t} + \tau_{SLT}$$

$$\tau_{SLT} = \gamma_0 \tau_d \boldsymbol{m} \times (\boldsymbol{m} \times \boldsymbol{\sigma})$$

Where $\boldsymbol{m} = \frac{\boldsymbol{M}}{M_s}$ is the unit magnetization vector with $M_s$ the saturation magnetization, $\boldsymbol{H}_{eff}$ is an effective magnetic field including an effective perpendicular anisotropy field $\boldsymbol{H}_{K,eff}$ ($= H_K - N_d M_s$), an external exchange bias field $H_X$, and the interfacial anisotropic exchange field due to the Dzyaloshinskii-Moriya interaction (DMI), $\gamma_0$ is the gyromagnetic ratio, $\alpha$ is the Gilbert damping parameter, $\tau_{SLT}$ is the SLT term, $\tau_d = \frac{\hbar \theta_{SH}^{eff} J_c}{2e\mu_0 M_s t_F}$ is the magnitude of SLT, and $\boldsymbol{\sigma}$ is the direction of spin polarization. The polarity of SOT torque is determined by the direction of the in-plane current $J_c$.

PtMn is chosen as the AFM layer due to its higher spin Hall angle compared to other CuAu-I-type AFM [29-31] such as IrMn, PdMn, and FeMn. The spin Hall angle $\theta_{SH}^{eff}$ of PtMn varies from 0.06 up to 0.24 in different studies [29-32]. Herein, we assume $\theta_{SH}^{eff} = 0.12$. The PMA is achieved only in a limited range of CoFeB layer thickness; therefore, we fix the thickness at 1 nm which is within the range. The structure of a Hall bar stack is assumed, from the substrate side, Ta(5 nm)/PtMn(9 nm)/CoFeB(1 nm)/MgO(1.6 nm)/Ta(2 nm). The saturation magnetization of CoFeB layer is $M_S = 1.2 \times 10^6 \ A/m$ [9]. For this sufficiently thin CoFeB layer, the PtMn (AFM) / CoFeB (FM) / MgO structure induces a strong perpendicular easy axis with an effective anisotropy field $H_K^{eff}$ between 120 mT and 500 mT along Z-axis [12, 13, 33-35] and the anisotropy field $H_K$ is between 1.5 T and 2 T [36]. The exchange bias field $H_X$ is assumed to be 15 mT along X-axis, which can be achieved by annealing at 300 °C under an in-plane magnetic field of 1.2 T along the Y-axis for 2 hours [13] (Lau et al [11] reached an exchange bias up to 50 mT by replacing MgO with Ru).

Table 1. Simulation Parameters of SOT Switching of CoFeB Nanopillar

| Parameter | Description | Values [a)] |
|---|---|---|
| **Nanopillar Dimensions** | Diameter × Thickness | 112 nm × 1 nm |
| | | (14 nm × 1 nm) |
| **Cell size** | Length × Width × Thickness | 1 nm × 1 nm × 1 nm |
| $\gamma_0$ | Gyromagnetic ratio | $2.211 \times 10^5 \ m/A \cdot s$ |
| $\alpha$ | Gilbert damping factor | 0.02 |
| A | Exchange constant | $20 \times 10^{-12} \ J/m$ |
| $M_S$ | Saturation magnetization | $1.2 \times 10^6 \ A/m$ |
| $J_C$ | Charge current density | $2.2 \times 10^{12} \ A/m^2$ |
| | | ($2.5 \times 10^{12} \ A/m^2$) |
| $\theta_{SH}^{eff}$ | Spin Hall angle | 0.12 |
| $H_K$ | Perpendicular anisotropy field | 1.5 T |
| $H_X$ | Exchange bias field | 15 mT |
| DMI | Dzyaloshinskii-Moriya interaction factor | $0.5 \ mJ/m^2$ |

[a)] The parameters used in the simulation on SOT switching of 14 nm nanopillar are listed in the parentheses. Smaller nanopillar requires larger charge current density $J_c$ to flip its magnetization, and the switching process generates smaller emf.

## 4. Results

## 4.1 SOT Switching of 14 nm FM Nanopillar

Herein, we conducted the OOMMF simulations on the CIP geometry AFM/FM nanostructure mentioned in the foregoing sections. In-plane charge current pulses are applied through the AFM layer. The spin Hall torque is directed along $\boldsymbol{m} \times (\boldsymbol{m} \times \boldsymbol{\sigma})$. The magnetization dynamics of the FM nanopillar with diameter of 14 nm and thickness $t_F = 1\ nm$ (denoted as FM14-SOTNS in this paper) are monitored. Figure 3(a) – (f) show the time-dependent change of the charge current pulses, the Z-component magnetization of the FM14-SOTNS structure, and the calculated electric field on the surface of FM14-SOTNS. Electric fields are generated during the magnetization switching process. Significant precessional motions are observed when the charge current switches direction or turns on/off, which is because the magnetization direction deviates from its equilibrium direction (i.e., the perpendicular easy axis). The precessional motions induce high-frequency electric field with magnitudes between 0 and 70 mV/mm as shown in Figure 3(e). Figure Figure 3(f) is an enlarged view of one electric field stimulus pulse with pulse width of ~2 ns and an intensity of 5 – 70 mV/mm. The stimulus frequency of electric field pulses can be controlled by setting the charge current switching frequency between 60 Hz and 200 Hz.

The magnitude distribution of Z-component of magnetic stray field, $H_z$, from the surface of FM14-SOTNS is recorded in Figure 3(g). Figure 3(h) shows the magnetic stray field decays with the distance from nanopillar surface to a distance of 100 nm along Z direction. The trend line shows that $H_z$ drops to 1 A/m at a distance of around 300 nm. Figures 3(i – p) are the magnetization evolutions of FM14-SOTNS after applying the charge current pulse. Each arrow represents an area of 1 nm × 1 nm. Magnetization in the FM14-SOTNS flips from +Z direction to -Z direction within 2 ns. Switching is completed by involving SOT and DMI driven domain wall motion. The parameters used in this OOMMF simulation are listed in Table 1.

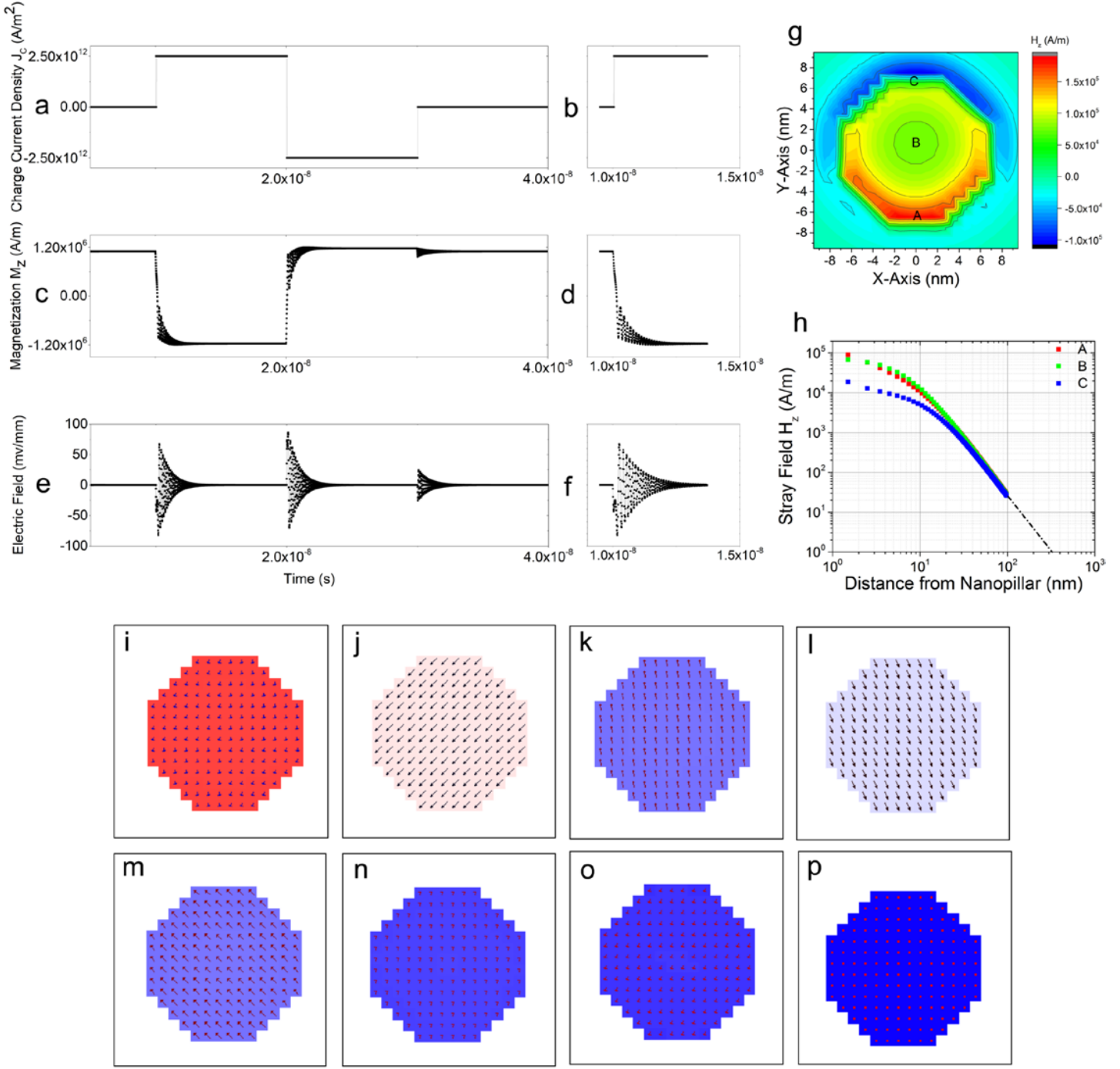

Figure 3. (a) Applied charge current pulses. $|J_C| = 2.5 \times 10^{12} \ A/m^2$, $+/-$ signs indicate $J_C$ flows along $+/- X$ axis. (b) Charge current $+J_C$ is on at 10 ns. (c) Magnetization component $|M_z| \sim 1.2 \times 10^6 \ A/m$, $+/-$ signs indicate $M_z$ points to $+/- Z$ direction. (d) Flipping $M_z$ from +Z direction to -Z direction by a charge current pulse $+J_C$ within 3 ns. (e) Calculated real-time electric field based on Maxwell-Faraday law. (f) Real-time electric field within 3 ns of charge current pulse $+J_C$. (g) Magnetic stray field component $H_z$ on the top surface of 14 nm FM nanopillar when the magnetization is fully switched ($M_z/M_s = 1$). Points A and C are at the edge of nanopillar, and point B is in the center of nanopillar. (h) Magnetic stray field component $H_z$ attenuates from top surface of nanopillar to a distance of 100 nm from surface along Z-Axis. (i) to (p) are the simulated magnetization evolution in a 14 nm diameter FM nanopillar at △t=0, 0.05 ns, 0.13 ns, 0.20 ns, 0.28 ns, 0.90 ns, 1.21 ns and 2.06 ns after applying the charge current pulse.

**4.2 SOT Switching of 112 nm FM Nanopillar**

In the foregoing section, we discussed the SOT-based magnetization switching behavior in a FM14-SOTNS structure with a diameter of 14 nm, where it is in the single domain regime. We observed that the switching behavior in the device can be described by one macrospin. In this section, we increased the diameter of FM nanopillar to 112 nm while keeping the thickness 1 nm (denoted as FM112-SOTNS in this paper), entering the nucleation dominated regime. The time evolutions of charge current, Z-component magnetization in FM nanopillar, and calculated electric field are plotted in Figure 4(a – f). During one magnetization switch cycle (from +Z to –Z direction or vice versa), the precessional motions of the magnetization in FM nanopillar induce high frequency electric field with magnitudes between 0 and 200 mV/mm as shown in Figure 4(e). Figure 4(f) is an enlarged view of one electric field stimulus pulse with pulse width of ~3 ns and an intensity of 5 – 200 mV/mm.

The Z-component stray field pattern and the stray field decay rate along the Z direction distances are plotted in Figure 4(g) & (h). The trend line shows that $H_z$ drops to 1 A/m at a distance of around 6 μm. In addition, as the FM nanopillar diameter increases from 14 to 112 nm, the overall time for a complete magnetization flip (from +Z to –Z direction or vice versa) increases from 2 ns to 3 ns. Figure 4 (i – r) shows the magnetization evolution of FM112-SOTNS structure vs time. Each arrow represents an area of 7 nm × 7 nm. Magnetization in 112 nm FM nanopillar flips from +Z direction to -Z direction within 3 ns. Switching is completed by involving SOT and DMI driven domain wall motion. The parameters used in this simulation are listed in Table 1.

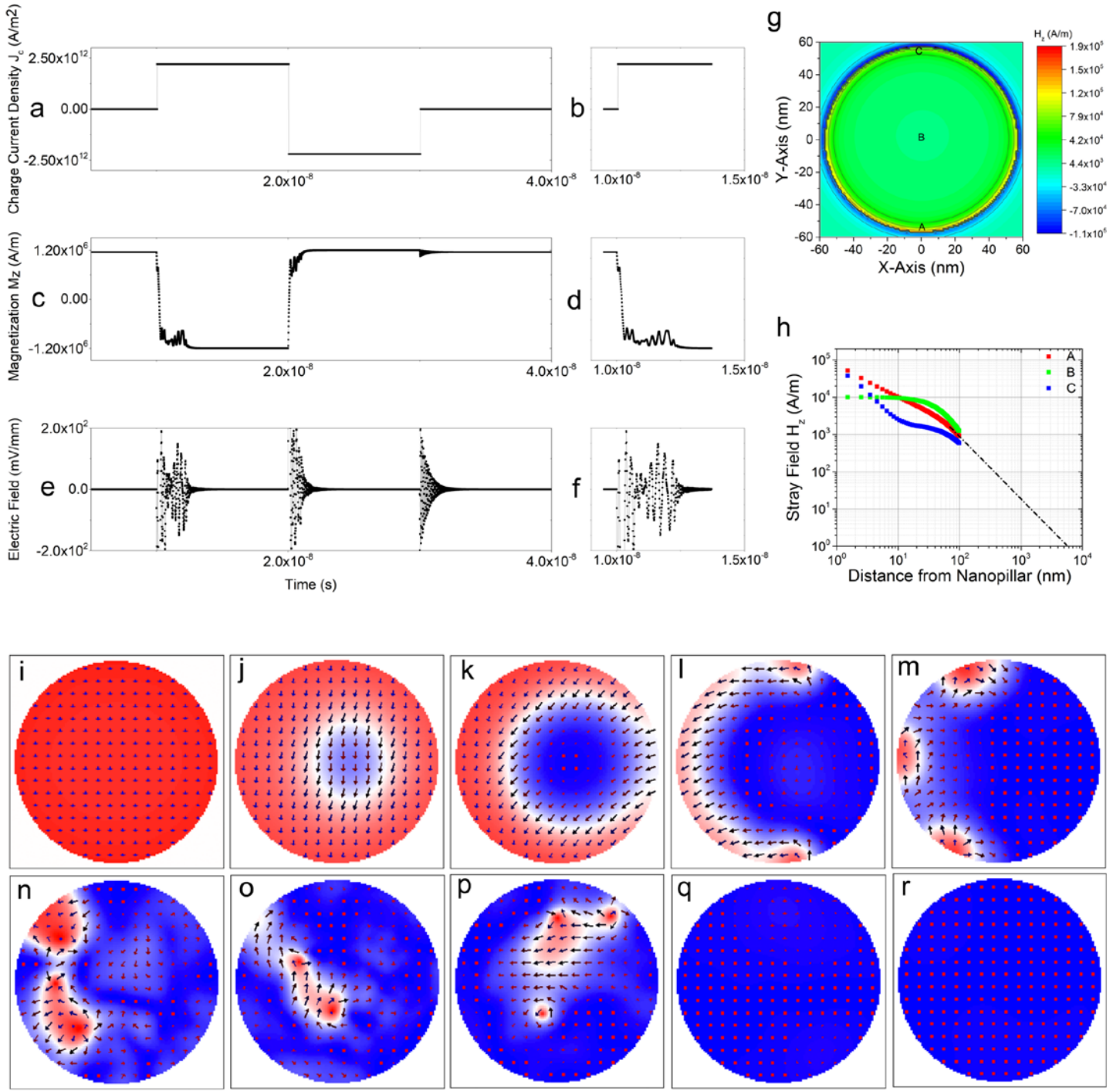

Figure 4. (a) Applied charge current pulses. $|J_C| = 2.2 \times 10^{12}\ A/m^2$, $+/-$ signs indicate $J_C$ flows along $+/-X$ axis. (b) Charge current $+J_C$ is on at 10 ns. (c) Magnetization component $|M_z|{\sim}1.2 \times 10^6\ A/m$, $+/-$ signs indicate $M_z$ points to $+/-Z$ direction. (d) Flipping $M_z$ from +Z direction to -Z direction by a charge current pulse $+J_C$ within 3 ns. (e) Calculated real-time electric field based on Maxwell-Faraday law. (f) Real-time electric field within 3 ns of charge current pulse $+J_C$. (g) Magnetic stray field component $H_z$ on the top surface of 112 nm FM nanopillar when the magnetization is fully switched ($M_z/M_s = 1$). Points A and C are at the edge of nanopillar, and point B is in the center of nanopillar. (h) Magnetic stray field component $H_z$ attenuates from top surface of nanopillar to a distance of 100 nm from surface along Z-Axis. The trend line shows that $H_z$ drops to 1 A/m at a distance of around 5 μm. (i) to (r) are the simulated magnetization evolution in a 112 nm diameter FM nanopillar at Δt=0, 0.12 ns, 0.20 ns, 0.27 ns, 0.35 ns, 0.42 ns, 0.50 ns, 0.65 ns, 2.79 ns and 2.94 ns after applying the charge current pulse.

## 5. Conclusions and Future Perspectives

To summarize, we have initially explored the feasibility of applying magnetic spintronic nanodevices for neuromodulation applications. An AFM/FM structure is employed where the AFM layer supplies an in-plane exchange bias field on the adjacent FM layer and meanwhile generates SOT to enable deterministic switching of perpendicular magnetization. This current driven magnetization dynamics in SOTNS has no mechanically moving parts. Furthermore, the size of SOTNS can be down to tens of nanometers. Thus, arrays of SOTNSs could be fabricated, integrated together and patterned on a flexible substrate, which gives us much more flexible control of the neuromodulation. The magnetization dynamics of two AFM/FM structures were simulated using OOMMF: FM14-SOTNS and FM112-SOTNS, which are 1 nm thick FM nanopillars with diameters of 14 nm and 112 nm. The time evolutions of charge current, Z-component magnetization in FM nanopillars, and calculated electric fields are plotted. During one magnetization switch cycle (from +Z to –Z direction or vice versa), the precessional motions of the magnetization in FM nanopillar induce high frequency electric fields. As the FM nanopillar diameter increases from 14 nm to 112 nm, the magnitudes of induced electric fields increase from tens of mV/mm to hundreds of mV/mm. In addition, the time duration of electric field in each neuromodulation cycle increases from 2 ns to 3 ns.

The fields of neuroscience and neural engineering have seen rapid growth during the last few decades, especially in new fabrication and materials technologies to produce high density, miniaturized, and customized electrode arrays for sensing and stimulating neurons [37-39]. Nanometer-scale devices exploiting spintronics can be a key technology in this context. Spintronic nanodevices offer a plethora of novel mechanisms which can be harnessed into new device paradigms with the potential to drive progress in the sensing and modulation of neuron activities. Other features of spintronic nanodevices for neuromodulations include tunable magnetic dynamics, low power consumption, no wear out (no mechanical moving parts) during function cycles, and their magnetic/electric performances are not affected by the encapsulation of astrocytes or any other cells. In addition, over the past decade, there have been exciting developments in using polymer coatings (e.g. PEDOT), hydrogels, and surface modifications to improve the tissue and functional interaction of the nanodevices with the surrounding neurons. It can be foreseen that spintronic nanodevices along with neural engineering will enable more precise diagnostics and therapeutics for brain disorders.

**Acknowledgements**

This study was financially supported by the Institute of Engineering in Medicine of the University of Minnesota, National Science Foundation MRSEC facility program, the Distinguished McKnight University Professorship, Centennial Chair Professorship, Robert F Hartmann Endowed Chair, and UROP program from the University of Minnesota.

**Conflict of Interest**

The authors declare no conflict of interest.